\documentclass[11pt,letterpaper]{article}
\usepackage[utf8]{inputenc}

\usepackage[includeheadfoot,
            marginratio={1:1,2:3},
            width=412pt,
            height=688pt,]{geometry}

\usepackage{amsmath}
\usepackage{amsfonts}
\usepackage{amssymb}
\usepackage{stmaryrd}
\usepackage{graphicx}
\usepackage{mathtools}
\usepackage{amsmath}
\usepackage{fontenc}
\usepackage{mathtext}

\usepackage{empheq}
\usepackage{paralist}

\usepackage{amssymb}
\usepackage{color}
\makeatletter
\renewcommand*\env@matrix[1][*\c@MaxMatrixCols c]{%
  \hskip -\arraycolsep
  \let\@ifnextchar\new@ifnextchar
  \array{#1}}
\makeatother

\newcommand{\nc}{\newcommand}
\nc{\lb}{\llbracket}
\nc{\rb}{\rrbracket}
\nc{\gl}{\llbracket}
\nc{\gr}{\rrbracket}

\newcommand{\eq}[1]{\begin{equation}
                     \begin{split} #1 \end{split}
                     \end{equation}}

\allowdisplaybreaks[2]
\numberwithin{equation}{section}

\begin{document}

\vspace*{-1.5cm}
\begin{flushright}
 {MPP-2014-391}
\end{flushright}

\vspace{1.5cm}
\begin{center}
{\bf\Large T-duality, Non-geometry  and Lie Algebroids \\[0.2cm]
in Heterotic Double Field Theory
}
\vspace{0.4cm}

\end{center}

\vspace{0.35cm}
\begin{center}
  Ralph Blumenhagen and  Rui Sun
\end{center}

\vspace{0.1cm}
\begin{center}
\emph{Max-Planck-Institut f\"ur Physik (Werner-Heisenberg-Institut), \\
   F\"ohringer Ring 6,  80805 M\"unchen, Germany } \\[0.1cm]
\vspace{0.25cm}

\end{center}

\vspace{1cm}


\begin{abstract}
A number of  issues in heterotic double field theory are
studied. This includes the analysis of the T-dual configurations
of a flat constant gauge flux background, which turn
out to be  non-geometric.
Performing  a field redefinition to a  non-geometric frame, these
T-duals take a very simple form reminiscent of the constant
$Q$- and $R$-flux backgrounds.
In addition, it is shown how the analysis of arXiv:1304.2784
generalizes to heterotic generalized geometry. For every
field redefinition specified by an $O(D,D+n)$ transformation,
the structure of the resulting supergravity  action is governed  by the differential
geometry of a corresponding Lie algebroid.
\end{abstract}

\clearpage


\section{Introduction}

The description and understanding of non-geometric string backgrounds
have been under investigation during the last years. As with many
other developments in string theory, the exploration of the
consequences of T-duality has been a good guide in this respect
\cite{Dabholkar:2002sy,Hull:2004in,Shelton:2005cf,Dabholkar:2005ve}.
The classic example is to perform successive  T-dualities via
the Buscher rules \cite{Buscher:1987sk,Buscher:1987qj} applied
to a flat
closed string background with constant $H$-flux \cite{Shelton:2005cf}.
This led to the chain of flux backgrounds with fluxes
$H_{abc} \to F^a{}_{bc} \to Q_c{}^{ab} \to R^{abc}$
where it was shown that the $Q$-flux background is non-geometric
globally and the $R$-flux background even locally. It was shown
that these non-geometric backgrounds take a very simple form, when
expressed not in the geometric frame $(g_{ij},B_{ij})$ but in a
so-called non-geometric frame $(\tilde g_{ij},\beta^{ij})$, where the new metric
and the bi-vector are related to the geometric frame via a field redefinition.

In order to properly describe such backgrounds, one needs to go beyond
the usual effective supergravity description of string theory. In this
respect,  two approaches were followed. The first one is  generalized
geometry \cite{Hitchin:2004ut,Gualtieri:2003dx,Grana:2008yw,Coimbra:2011nw},
where one extends the tangent bundle of a manifold such that
diffeomorphisms
and $B$-field gauge transformations can be described in a single
geometric framework. Concretely, the metric and the Kalb-Ramond field are unified
in a generalized metric on the bundle $T\oplus T^*$.
A more ambitious approach is to develop
a theory which is manifestly invariant under T-duality. This led to
double field theory (DFT), where not only the bundle but also the
coordinates themselves are extended to a doubled space by introducing
winding coordinates.
A first approach followed a frame-like formulation
\cite{Siegel:1993xq,Siegel:1993th} which was further worked out  in
\cite{Hohm:2010xe,Geissbuhler:2013uka}. Later, using string field
theory,   an equivalent  generalized metric formulation
was found \cite{Hull:2009mi,Hohm:2010jy,Hohm:2010pp}.
However,  DFT not only features a global $O(D,D)$ symmetry
but also  the local symmetries, due to the winding dependence, are enhanced.
For recent reviews of DFT see \cite{Aldazabal:2013sca,Berman:2013eva, Hohm:2013bwa}.

Whether generalized geometry allows for a description of non-geometric
backgrounds has been investigated in a series of papers.
In particular, in \cite{Andriot:2011uh,Andriot:2012an,Andriot:2013xca} the question has been asked
what form the usual supergravity action takes in the non-geometric frame
variables and whether this action might already  be  sufficient for
the global description of non-geometric backgrounds. This led to the definition
of so-called $\beta$-supergravity.
In \cite{Blumenhagen:2012nt,Blumenhagen:2013aia} the general structure
of  such $O(D,D)$ induced field redefinitions
was clarified in the framework of generalized geometry. The two main
results were that for every such field redefinition, one can associate a
corresponding Lie algebroid so that the redefined supergravity action
is governed by the differential geometry of that Lie algebroid.
It turned out that in each patch this provides are good description of
the background, but that the transition functions needed for the
global description in general are not part of the local symmetries of generalized
geometry\footnote{For possible exceptions see \cite{Andriot:2014uda}.}.

On the contrary, due to the existence of extra local symmetries in DFT, i.e. the generalized
diffeomorphisms, the latter  admits  a global description of the
$Q$ and $R$-flux backgrounds. The non-geometry shows up
for the $Q$-flux background in a winding dependence of the
transition function between two patches and for the  $R$-flux in an
explicit winding dependence of the background field itself.

A natural generalization of bosonic DFT is heterotic DFT
\cite{Siegel:1993xq,Siegel:1993th, ReidEdwards:2008rd, Hohm:2011ex}, where the
latter also includes the gauge fields present in the heterotic string.
In generalized geometry the heterotic string was also discussed
in \cite{Andriot:2011iw, Garcia-Fernandez:2013gja,Baraglia:2013wua}(see also
\cite{Anderson:2014xha, delaOssa:2014cia}).
For abelian gauge fields this
generalization is formally straightforward extending the global
symmetry group from $O(D,D)$ to $O(D,D+n)$.  For every gauge field
$A^\alpha$ a new coordinate $y^\alpha$  is introduced, thus extending also the generalized metric
so that it includes the gauge fields. The main relations of DFT
remain unchanged so that  the action still has the same form
as for bosonic DFT, but just for the extended generalized metric.
This abelian heterotic DFT can be gauged which also allows the
description of non-abelian gauge groups \cite{Hohm:2011ex,Grana:2012rr}. However, in this process the
part of the global symmetry group is broken to $O(D,D)$.

It was observed in \cite{Bedoya:2014pma} that, in contrast to bosonic DFT, the
action of T-duality gives the Buscher rules including $\alpha'$ corrections.
In the same work, a suggestion has been made how heterotic DFT can be
further generalized to also accommodate the leading order gravitational
$\alpha'$ corrections, including e.g. the well known Chern-Simons terms
involving the spin-connection.  There has been quite some interest
recently
on how to incorporate   such $\alpha'$ corrections in the framework of
generalized geometry \cite{Coimbra:2014qaa, delaOssa:2014msa} and DFT
\cite{Bedoya:2014pma,Hohm:2013jaa,Hohm:2014eba,Hohm:2014xsa}.

In this paper, in some sense, we  take a step back from these more formal
developments and investigate some comparably simple questions which,
as  we think,  are nevertheless important to clarify.
For instance, to our knowledge, it is not clear what the heterotic
T-dual of a constant gauge flux background is. The same question for
the S-dual background of a type I string led to the discovery
of $D$-branes and $O$-planes. Indeed the T-dual of a $D9$-brane
carrying a constant gauge flux in type I is the type I' string with
a $D8$ brane intersecting the $O8$-plane at an angles.
In the heterotic case there are no $8$-branes so what is the T-dual?
Not unrelated, one can ask whether for heterotic DFT, there exist
an analogous chain of T-dual fluxes as for bosonic DFT. Of course,
the $H_{ijk}\to F_{ij}{}^k\to Q_i{}^{jk}\to R^{ijk}$ chain will still exist,
but what about a similar chain starting
with an abelian  constant gauge flux $G_{ij} \to \ldots$? Does it also give rise
to new types of non-geometric fluxes? Clearly, these are questions one
can now approach in the framework of heterotic DFT.

We will find that indeed after one T-duality, one gets a non-geometric
gauge flux background
that is in many ways analogous to the $Q$-flux background. It is
locally still geometric and the non-geometry appears in the transition
functions in the sense that there appears a dependence on a winding
coordinate.
 Moreover, also for heterotic DFT one can perform a
field redefinition  to a  non-geometric frame in which the fundamental
fields are a dual  metric $\tilde g_{ij}$, a bi-vector $\beta^{ij}$ and a gauge
one-vector $\tilde A^i$. We will see that one indeed gets a chain of
fluxes  $G_{ij}\to J^i{}_j \to \tilde G^{ij}$, where the latter two are
non-geometric.
One can trace back that, in this case, the non-geometry arises due to the $\alpha'$ corrections to the
T-duality rules \cite{Bergshoeff:1994dg,Serone:2005ge}.

Having realized that a field redefinition can be important for the
description of non-geometric backgrounds, we can ask how the analysis
of \cite{Blumenhagen:2013aia} generalizes to the heterotic case. Can the effect of an $O(D,D+n)$
induced field redefinition still be described by the differential
geometry of a corresponding Lie algebroid? We will see that this is
indeed possible and explicitly present the corrections due to
the existence of the gauge field.  As for the original version,
the local symmetries of the redefined action are only the redefined versions of
diffeomorphisms,  $B$-field and $A$-field gauge transformation. This  implies that
a single such action cannot globally describe non-geometric
backgrounds, which need winding coordinates
to appear either in the transition functions ($Q$-flux) or in the
background  itself ($R$-flux).

This paper is organized as follows:
In section $2$ we briefly introduce the setup for heterotic DFT and
the basics we need for our discussion. In section $3$ we will perform
in detail successive T-dualities on a toroidal constant gauge flux background
 and show how a field redefinition to a non-geometric frame simplifies
 the description of the T-dual backgrounds.
We derive explicitly the form of the relevant heterotic fluxes and
comment on the consequences for a potential non-associativity and
for S-duality to the type I string.
In section $4$ we  discuss the $O(D,D+n)$ induced  field redefinitions
and identify the corresponding Lie algebroid. In particular, we
present the form of the action in the previously introduced non-geometric frame.

\section{Review of heterotic DFT}

In this section we briefly review the bosonic sector of  heterotic
DFT, where we focus on those features which are important in the
remainder of this note.
The bosonic sector of heterotic DFT with abelian gauge fields is a straightforward
generalization of bosonic DFT \cite{Hohm:2011ex}. This is expected, as
abelian gauge fields appear by dimensional reduction of gravity theories.

The low-energy effective action of the massless bosonic sector for the
heterotic string is described by the action
\eq{\label{effaction}
{\cal S}=\int dx \sqrt{g} \, e^{-2\phi}\, \Big(R+4(\partial
\phi)^2-\frac{1}{12}H^{ijk}{H}_{ijk}-\frac{1}{4}{G}^{ij}{}_\alpha
 G_{ij}{}^{\alpha}\Big)\, ,
}
in which the field strength of the non-abelian gauge fields is defined as
\eq{
\label{fieldstrengthG}
G_{ij}{}^{\alpha}=\partial_i A_j{}^\alpha-\partial_j A_i{}^\alpha+g_0 \,[A_i,
A_j]^\alpha
}
and the strength of the Kalb-Ramond field is modified by the Chern-Simons three-form,
\eq{
\label{hetthreeformH}
H_{ijk}=3\Big(\partial_{[i}
B_{jk]}-\kappa_{\alpha\beta} A_{[i}{}^\alpha \partial_j A_{k]}{}^\beta
-\frac{1}{3}g_0\, \kappa_{\alpha\beta} \,A_{[i}{}^\alpha [A_j,A_{k]}]^\beta\Big)\, .
}
Here $\kappa_{\alpha\beta}$ denotes  the Cartan-Killing metric of the gauge group. In the abelian
case, this is simply the unit matrix,
$\kappa_{\alpha\beta}=\delta_{\alpha\beta}$.
Note that the order in $\alpha'$ can be made visible by scaling $A_i{}^\alpha\to
\sqrt{\alpha'} A_i{}^\alpha$.
From now on we consider abelian gauge fields.  Moreover, the higher
derivative correction of $H$ due to the gravitational Chern-Simons form
will be not considered throughout this paper.

\subsection{The generalized metric}

In the DFT formulation of the abelian heterotic sting \cite{Hohm:2011ex}, for each gauge
field $A^\alpha$ one introduces a new coordinate $y^\alpha$ so that
the entire DFT lives on a $2D+n$ dimensional space with coordinates
\eq{ X^M = (\tilde x_i, x^i, y^\alpha)\, .
}
The global symmetry group is enhanced from $O(D,D)$ to $O(D,D+n)$, the
T-duality group of the heterotic string.
The doubled coordinates $X^M$ transform as an $O(D,D+n)$ vector
\eq{X^{'M}=h^M{}_N\, X^N\, ,\qquad  h\in O(D,D+n)\, .
}
As in bosonic DFT, one introduces an $O(D,D+n)$ invariant metric
\eq{\label{eta}
\eta_{MN} =
 \begin{pmatrix}
  0 & \delta^{i}{}_{j} & 0 \\
  \delta_{i}{}^{j} & 0 & 0 \\
  0  & 0  & \delta_{\alpha \beta}
 \end{pmatrix}
}
satisfying
\eq{\eta^{MN}=h^M{}_P\,  h^N{}_Q\; \eta^{PQ}\, .
}
This $O(D,D+n)$  metric is used to pull up and down capital indices
like $M$. Accordingly, the generalized derivatives and gauge
parameters are given as
\eq{\partial_M = (\tilde\partial^i, \partial_i, \partial_\alpha), \quad   \xi^M
= (\tilde\xi_i,  \xi^i, \Lambda^\alpha)\, .
}
As shown in \cite{Hohm:2011ex}, one can proceed along the lines of bosonic DFT
and  introduce a generalized Lie derivative and a C-bracket. Then,
closure of the algebra is guaranteed, if one introduces the heterotic
strong constraint
\eq{
\partial_M f \, \partial^M g = \tilde\partial^i f \, \partial_i
g+\partial_i
f\, \tilde\partial^i g+\partial_\alpha f \, \partial^\alpha g=0
}
where $f$ and $g$  are arbitrary fields and gauge parameters.
This means that the heterotic level-matching condition
\eq{
\partial_M \partial^M f = 2\tilde\partial^i\partial_i
f+\partial_\alpha\partial^\alpha f=0
}
also has to hold for products of fields
and implies that locally there exist an $O(D,D+n)$ transformation rotating
the coordinates into a frame in which the fields only depend on the normal
coordinates $x^i$.

The heterotic DFT action can be expressed  in terms of a  generalized metric and an $O(D,D+n)$ invariant dilaton $d$ defined by
$e^{-2d}=\sqrt{g}e^{-2\phi}$.
The metric ${\cal H}^{MN}$ transforms covariantly under $O(D,D+n)$
\eq{{\cal H}^{'MN}(X^{'})=h^M{}_P\,  h^N{}_Q\, {\cal H}^{PQ}(X)
}
and is parameterized in terms of the metric $g_{ij}$, the Kalb-Ramond
field $B_{ij}$ and the gauge fields $A_i{}^\alpha$ as
\eq{
\label{genmetric}
{\cal H}_{MN} =
 \begin{pmatrix}
  g^{ij} & -g^{ik}C_{kj} & -g^{ik}A_{k\beta} \\[0.1cm]
  -g^{jk}C_{ki} & g_{ij}+C_{ki}g^{kl}C_{lj}+A_i{}^\gamma A_{j\gamma} &
C_{ki}g^{kl}A_{l\beta}+A_{i\beta} \\[0.1cm]
  -g^{jk}A_{k\alpha }  & C_{kj}g^{kl}A_{l\alpha}+A_{j\alpha}  &
\delta_{\alpha\beta}+A_{k\alpha}g^{kl}A_{l\beta}
 \end{pmatrix}
}
where only the combination
\eq{C_{ij}=B_{ij}+\frac{1}{2}A_i{}^{\alpha} A_{j \alpha}
}
appears. Note that $C_{ij}$  splits into a symmetric and an antisymmetric part as
\eq{
C_{(ij)}=\frac{1}{2} A_i{}^\alpha A_{j \alpha }\, ,\qquad
C_{[ij]}=B_{ij}\, .
}
Written in terms of the generalized metric \eqref{genmetric}, the form
of the heterotic DFT action
is identical to the action of bosonic DFT.

\subsection{Generalized vielbeins in heterotic DFT}

In analogy to  bosonic DFT, one can also introduce a generalized vielbein
$E^A{}_{M}$ so that
 \eq{
{\cal H}_{MN} =E^A{}_{M}\, S_{AB}\, E^B{}_N
}
with the constant generalized metric
\eq{
S_{AB} =
 \begin{pmatrix}
  s^{ab} & 0 & 0 \\
  0 & s_{ab} & 0 \\
  0 & 0  & s_{\alpha\beta}
 \end{pmatrix}
}
and $s_{ab}={\rm diag}(-, +,\ldots,+), s_{\alpha\beta}={\rm diag}(+,\ldots,+)$.
One finds\footnote{Note that compared to \cite{Geissbuhler:2013uka}, we have two
different signs  in the definition of the vielbein.  This is because
we want to be consistent with the generalized metric as defined in \cite{Hohm:2011ex}.}
\eq{
\label{vielbein}
E^{A}{}_{M} =
 \begin{pmatrix}
  e_a{}^i & -e_a{}^k C_{ki} & -e_a{}^{k}A_{k\beta} \\[0.1cm]
  0& e^a{}_{i}      & 0 \\[0.1cm]
  0& A_i{}^{\alpha}{}   & \delta^{\alpha}{}_{\beta}
  \end{pmatrix}\, ,\quad
E_{A}{}^{M} =
 \begin{pmatrix}
  e^a{}_i & 0 & 0\\[0.1cm]
-e_a{}^k C_{ki} & e_a{}^i & -e_a{}^{k} A_{k}{}^{\beta} \\[0.1cm]
   A_{i\,\alpha}   & 0 &  \delta_{\alpha}{}^{\beta}
  \end{pmatrix}
}
which also satisfies
\eq{\eta_{MN}=E^A{}_M\, E_{AN}\, .
}
Now one defines the generalized derivative as
\eq{
D_{A}=E_{A}{}^{M} D_M= (\tilde D^a, D_a, D_{\alpha})
}
leading in components to
\eq{
\tilde D^a &= \tilde \partial^a \\
           D_a &=\partial_a - B_{ai}\,\tilde\partial^i
-\frac{1}{2}A_a{}^\alpha A_{i\alpha} \,\tilde\partial^{i}-A_a{}^\gamma\,
\partial_{\gamma}\\
     D_{\alpha} &=\partial_{\alpha}+A_{i\alpha}
     \,\tilde\partial^{i}\, .
}
Introducing the generalized Weitzenb\"ock connection as
\eq{\Omega_{ABC}&=D_A E_B{}^N E_{CN}
}
the generalized fluxes of heterotic DFT are defined as
\eq{{\cal F}_{ABC}=&E_{CM} {\cal L}_{E_A}
E_B{}^M=\Omega_{ABC}+\Omega_{CAB}-\Omega_{BAC}\, .
}
In a  holonomic basis one finds e.g. the three-form flux
\eq{
\label{holHflux}
H_{ijk}=-3\Big(\partial_{[\underline
i}B_{\underline{jk}]}-\delta_{\alpha\beta} A_{[\underline i}{}^\alpha
\partial_{\underline j} A_{ \underline k]}{}^\beta\Big)
}
which  is precisely the field strength of the Kalb-Ramond field modified by the
Chern-Simons three-form \eqref{hetthreeformH}.
In section \ref{sec_fluxes}, all these generalized fluxes will be evaluated more
explicitly.

\section{Non-geometric backgrounds of heterotic DFT}

In this section we will use the formalism of heterotic DFT to
determine the T-dual of a heterotic string compactified on a two-torus $T^2$
with a constant gauge flux turned on. This is analogous to the
configuration of $T^3$ with constant $H$ flux for bosonic string
theory \cite{Shelton:2005cf, Blumenhagen:2014sba}.
In the latter case, this was the prototype example to detect after two T-dualities
the possibility of a non-geometric $Q$-flux background. Applying a
third T-duality led to the conjecture for the existence of an $R$-flux
background. Applying these T-dualities in the framework of DFT, the
non-geometry shows up in  the appearance of winding coordinates
in the transitions functions for the $Q$-flux and in the background
itself for the $R$-flux. Thus, in this sense a $Q$-flux background is
locally geometric but not globally, whereas an $R$-flux background is
non-geometric even locally.

\subsection{T-duality of a constant gauge flux background}
\label{sec_tduali}

Recall that under  a global $h\in O(D,D+n)$ transformation the
coordinates and the generalized metric behave as
\eq{
H^{'}=h^t\, H\, h\, , \quad X^{'}=h\, X\, , \quad \partial
^{'}=(h^t)^{-1}\, \partial\, .
}
Now, we consider a torus $T^2$  with a flat metric
$g_{ij}=\delta_{ij}$, vanishing Kalb-Ramond field $B$ and a constant
abelian gauge flux $G_{ij}$.  For the corresponding single gauge
potential $A^{(1)}=A$ we choose
\eq{
 A_{1}=f\, y \, , \qquad  A_{2}=0\, .
}
This gives the field strength
\eq{
G_{12}=-(\partial_1 A_2-\partial_2 A_1)=f\, .
}
On the $2$-torus the coordinates are periodically identified by
$(x, y)\sim (x+2\pi, y)\sim (x,y+2\pi)$.
For the gauge field to be well defined globally, one needs a
non-trivial transition function between the two patches $P=[0, 2\pi)$ and $Q=(0, 2\pi]$.
In the patch $P$ we have $A^{(P)}_1=f\, y$ while in the patch $Q$  the
gauge field is
$A^{(Q)}_1=f (y- 2\pi) $. These two patches can be glued smoothly together
by a gauge transformation $A_1^{(Q)}=A^{(P)}_1+\partial_1  \lambda^{(PQ)}$ with
\eq{
\label{concretegaugetrafo}
\lambda^{(PQ)}=-2\pi f x\, .
}
The generalized metric for this background in patch $P$ takes the form
\eq{{\cal H}^{(P)}_{MN} =
 \begin{pmatrix}[cc|cc|c]
  1& 0&    -\frac{(f y)^2}{2}  & 0&  -(f y)\\[0.1cm]
  0& 1&    0&  0&    0\\[0.1cm]
  \hline
 & & & & \\[-0.3cm]
  -\frac{(f y)^2}{2}  & 0 & 1+(f y)^2+\frac{(f y)^4}{4}& 0&
(f y)+\frac{(f y)^3}{2}\\[0.1cm]
  0&  0 &0&  1 &0\\[0.1cm]
  \hline
& & & & \\[-0.3cm]
   -(f y) & 0 &(f y)+\frac{(f y)^3}{2}&0 &1+(f y)^2
 \end{pmatrix}\, .
}
The transition to patch $Q$ is given by conjugation with an
appropriate $O(D,D+n)$ matrix ${\cal T}_{(PQ)}$, i.e.
\eq{
{\cal H}^{(Q)}={\cal T}^T_{(PQ)} \,\,{\cal H}^{(P)}\,\, {\cal T}_{(PQ)}\,
}
which in our case takes the form
\eq{
{\cal T}_{(PQ)} =
 &\begin{pmatrix}[cc|cc|c]
  1&  0  &-\frac{(2\pi f)^2}{2}  &0  &2\pi f \\
  0&  1  &0  &0  &0 \\
   \hline
    0& 0&1&0 & 0\\
    0&  0 &0  &1  &0 \\
     \hline
        0&  0  &-2 \pi f  &0  &1
 \end{pmatrix}\, .
}
In analogy to generalized  geometry such a matrix might be called an ``$A$-transform''.
Note that this is consistent with the discussion in \cite{Andriot:2013xca}, where the
transition matrix was calculated via the  vielbeins in the two patches
as ${\cal T}_{(PQ)}=E^{-1}_{(P)}\, E_{(Q)} $.

Now, we apply a T-duality in the $x$-direction, which in heterotic DFT
can be implement by conjugation ${\cal H}'={\cal T}_{1}^T\, {\cal H}\, {\cal T}_{1}$
with the special $O(2,3)$ transformation
\eq{{\cal T}_1=
\begin{pmatrix}[cc|cc|c]
0 & 0  & 1 &0 &0 \\
0 & 1  & 0 &0 &0  \\
  \hline
1 & 0  & 0 &0 &0  \\
0 & 0  & 0 &1 &0  \\
  \hline
0 & 0  & 0 &0 & 1
 \end{pmatrix}\, .
}
The upper $4\times 4$ dimensional part of the metric is the same as the
T-duality transformation for bosonic DFT.
Thus, we obtain in patch $P$
\eq{
\label{metricafteroneT}
{{\cal H}'}^{(P)} =
 \begin{pmatrix}[cc|cc|c]
  1+(f y)^2+\frac{(f y)^4}{4}& 0&    -\frac{(f y)^2}{2}  & 0&  (f y)+\frac{(f y)^3}{2}\\[0.1cm]
  0& 1&    0&  0&    0\\[0.1cm]
  \hline
& & & & \\[-0.3cm]
  -\frac{(f y)^2}{2}  & 0 & 1& 0&-(f y)
\\[0.1cm]
  0&  0 &0&  1 &0\\[0.1cm]
  \hline
& & & & \\[-0.3cm]
   (f y)+\frac{(f y)^3}{2}&0 &-(f y) & 0 &1+(f y)^2
 \end{pmatrix}
}
from which one  can directly  read off the new metric, Kalb-Ramond
field  and the gauge field as
\eq{
{g'}^{(P)} =&
   \begin{pmatrix}
 \frac{1}{1+(f y)^2+\frac{(f y)^4}{4}}& 0\\[0.3cm]
 0&1
 \end{pmatrix}
\, ,\qquad
 {B'}^{(P)}=0
\, ,\qquad
  {A'}^{(P)}=
 \begin{pmatrix}
-\frac{(f y)}{1+\frac{(f y)^2}{2}}\\[0.3cm]
0
 \end{pmatrix}\, .
}
Note that after one T-duality
one still gets a metric and a gauge field, where, as in the $Q$-flux
background, there  appears a non-trivial functional dependence in the denominators.
Moreover, these results are consistent with the
$\alpha'$ corrected Buscher rules for a  T-duality along a single
direction for the heterotic string given in
\cite{Serone:2005ge}.  In fact, as shown in appendix \ref{app_b} these are
precisely the T-duality rules following from the heterotic  DFT construction.

The new transition matrix to patch $Q$ is given by
\eq{
\label{transPQ1T}
{\cal T}'_{(PQ)} ={\cal T}_1^{T} {\cal T}_{(PQ)} {\cal T}_1=
 &\begin{pmatrix}[cc|cc|c]
  1&  0  &0  &0  &0 \\
  0&  1  &0  &0  &0 \\
   \hline
    -\frac{(2 \pi f)^2}{2}& 0&1&0 & 2\pi f\\
    0&  0 &0  &1  &0 \\
     \hline
        -2 \pi f&  0  &0  &0  &1
 \end{pmatrix}
}
which is not any longer a usual  $A$-transform, i.e. a gauge
transformation. This observation and the appearance of  strange
denominators already indicates that we are dealing here rather with a
non-geometric background (like the $Q$-flux for bosonic DFT).

In analogy to  bosonic DFT, one can introduce a field redefinition so
that the generalized metric is parameterized by a new metric $\tilde
g_{ij}$, a bi-vector $\tilde C^{ij}$ and a (one-)vector $\tilde A^i$ as
\eq{
\label{genmetricnongeo}
{\cal H}_{MN} =
 \begin{pmatrix}
\tilde g^{ij}+\tilde C^{ki}\, \tilde g_{kl}\,\tilde C^{lj}+\tilde A^i{}_\gamma
\,\tilde A^{j\gamma} & -\tilde g_{jk}\,\tilde C^{ki}
&\tilde C^{ki}\, \tilde g_{kl}\,\tilde A^{l}{}_{\beta}+\tilde
A^i{}_{\beta} \\[0.1cm]
  -\tilde g_{ik}\,\tilde C^{kj} &  \tilde g_{ij} & -\tilde g_{ik}\,\tilde
A^k{}_{\beta} \\[0.1cm]
  \tilde C^{kj}\, \tilde g_{kl}\,\tilde A^{l}{}_{\alpha}+\tilde A^j{}_{\alpha}  &
-\tilde g_{jk}\,\tilde A^k{}_{\alpha } &
\delta_{\alpha\beta}+\tilde A^k{}_{\alpha}\, \tilde g_{kl}\,\tilde
A^l{}_{\beta}
 \end{pmatrix}\, .
}
Here $\tilde C^{ij}=\beta^{ij}+{1\over 2} \tilde A^i{}_\alpha\, \tilde
A^{j\alpha}$, where $\beta^{ij}$ is the antisymmetric bi-vector appearing also in
bosonic DFT.  The generalized vielbein reads in this case
\eq{
\label{vielbeinnongeo}
E^{A}{}_{M} =
 \begin{pmatrix}
  \tilde e_a{}^i & 0 & 0 \\
  -\tilde e^a{}_k \tilde C^{ki} & \tilde e^a{}_{i}  &-\tilde e^a{}_k \tilde A^{k}{}_\beta \\
  \tilde A^{i\alpha}& 0   & \delta ^{\alpha}{}_{\beta}
  \end{pmatrix}\,.
}
Comparing \eqref{metricafteroneT} with the form of the generalized metric in the
so-called non-geometric frame \eqref{genmetricnongeo}, one can read off
\eq{
{\tilde{g}}^{\prime(P)} =&
   \begin{pmatrix}
 1 & 0\\
 0&1
 \end{pmatrix}
\, ,\qquad
  {\tilde{A}}^{\prime (P)}=
 \begin{pmatrix}
 f y\\
0
 \end{pmatrix}\, ,
}
with $\beta^{ij}=0$.
This shows that in this frame the T-dual configuration takes a very
simple form. Moreover, using \eqref{transPQ1T} one can also find the metric
and the one-vector in patch Q
\eq{
{\tilde{g}}^{\prime(Q)} =&
   \begin{pmatrix}
 1 & 0\\
 0&1
 \end{pmatrix}
\, ,\qquad
  {\tilde{A}}^{\prime (Q)}=
 \begin{pmatrix}
 f (y-2\pi)\\
0
 \end{pmatrix}\, ,
}
Since the T-duality also changes  $x\to \tilde x$ in the gauge
transformation \eqref{concretegaugetrafo}, the ``gauge''
transformation connecting the one-vectors in patch $P$ and $Q$
becomes
\eq{
\label{concretegaugetrafong}
\tilde A^{\prime 1(Q)}=\tilde A^{\prime 1(P)}+\tilde\partial^1
\tilde\lambda^{(PQ)}\, \quad {\rm with}\quad
\tilde\lambda^{(PQ)}=-2\pi f \tilde x\, .
}
Note, that the transition function in this non-geometric frame
contains a winding coordinate so that indeed this T-dual background
is globally non-geometric, very similar to the $Q$-flux background for
bosonic DFT. The only difference is that the latter requires a
T-duality in two-directions in order to generate it from a constant
$H$-flux background. Finally, the new flux in this T-dual background
should be
\eq{
        J^1{}_2=-\partial_2 {\tilde A}^1=-f\, .
}

Applying to this configuration another  T-duality in the $y$ direction
only changes $y\to \tilde y$ in the generalized metric
\eqref{metricafteroneT}, so that in the non-geometric frame one
obtains
\eq{
{\tilde{g}}^{\prime\prime(P)} =&
   \begin{pmatrix}
 1 & 0\\
 0&1
 \end{pmatrix}
\, ,\qquad
  {\tilde{A}}^{\prime\prime (P)}=
 \begin{pmatrix}
 f \tilde y\\
0
 \end{pmatrix}\, ,
}
and similarly in patch $Q$. Therefore, like the $R$-flux background, this configuration is already
locally non-geometric, characterized by a non-geometric flux
\eq{
        \tilde G^{12}=-(\tilde\partial^1 {\tilde A}^2-\tilde\partial^2 {\tilde A}^1)=f\, .
}
Of course, at this stage the form of the new non-geometric fluxes
$J^i{}_j$ and $\tilde G^{ij}$ is just a guess. In the following
subsection, we derive the complete form of this new kind of fluxes
from the vielbein \eqref{vielbeinnongeo}.

\subsection{The fluxes of heterotic DFT}
\label{sec_fluxes}

In this section we derive the general form of the components of the
heterotic fluxes
\eq{
{\cal F}_{ABC}=&E_{CM} {\cal L}_{E_A}
E_B{}^M=\Omega_{ABC}+\Omega_{CAB}-\Omega_{BAC}\, .
}
In order to treat geometric and non-geometric components at the same time,
as in \cite{Geissbuhler:2013uka},
we use the general extended form of the generalized vielbein
\eq{
\label{fullvielbein}
E^{A}{}_{M} =
 \begin{pmatrix}
  e_a{}^i & -e_a{}^k C_{ki} & -e_a{}^{k}A_{k\beta} \\
  -e^a{}_k \tilde C^{ki} & e^a{}_{i}+e^a{}_{k}\tilde C^{kj}C_{ji} &-e^a{}_k \tilde A^{k}{}_\beta \\
  \tilde A^{i\alpha}& A_i{}^{\alpha}   & \delta^{\alpha}{}_{\beta}
  \end{pmatrix}
}
which combines \eqref{vielbein} and \eqref{vielbeinnongeo} into one
object. Recall that $\eta_{AB}=E_A{}^M E_{MB}$ implies that the
Weitzenb\"ock connection satisfies $\Omega_{ABC}=-\Omega_{ACB}$.
However, one can show that this relation ceases to be  satisfied with
the  full vielbein \eqref{fullvielbein}. Therefore, in the following we present the
geometric fluxes for the physically relevant case of $\tilde A^i{}_\alpha=0$
and the non-geometric fluxes for $A_i{}^\alpha=0$.
In addition, for simplicity  here we will work in a holonomic basis, the rather
lengthy generalizations to a non-holonomic basis can be found in
appendix \ref{app_fluxhet}.

The components of the derivatives  $D_{A}=E_{A}{}^{M} D_M $ become
\eq{
\label{derivcompb}
\tilde D^i=&\tilde \partial^i +\tilde C^{im}C_{mn} \tilde \partial^n
     -\tilde C^{im} \partial_m -\tilde A^{i\gamma}\partial_{\gamma}\\
D_i=&\partial_{i}-C_{im}\tilde\partial^m-A_{i}{}^\gamma\partial_{\gamma}\\
D_{\alpha} =&\partial_{\alpha}+A_{m\alpha}\tilde\partial^{m}+\tilde
          A^m{}_{\alpha}{} \partial_m\, .
}
For all three indices being of  normal or winding type we get the
fluxes $H,F,Q$ and $R$ including corrections depending on the gauge
fields $A$ and $\tilde A$.
In terms of the derivatives
\eqref{derivcompb}, for $\tilde A^i{}_{\alpha}=0$  the geometric  fluxes can be expressed as
\eq{
\label{HFgeo}
{\cal H}_{ijk}=&-3 D_{[\underline{i}} B_{\underline{jk}]}+3
D_{[\underline{i}}A_{\underline{j}\gamma} A_{\underline{k}]}{}^\gamma\\[0.1cm]
F^k{}_{ij}
=&-\tilde D^k B_{ij}+\tilde D^k A_{[\underline{i}\gamma}
A_{\underline{j}]}{}^\gamma
-2 D_{[\underline{i}}\beta^{km}
A_{\underline{j}]}{}_\gamma A_m{}^\gamma-2\beta^{km} D_{[\underline{i}}C_{m\underline{j}]}\, .
}
With $A_i{}^\alpha=0$ the non-geometric fluxes take the form
\eq{
\label{QRgeo}
Q_k{}^{ij}
=&-D_k \beta^{ij}+D_k\tilde A^{[\underline{i}\gamma}\tilde
A^{\underline{j}]}{}_\gamma
-\tilde C^{[\underline{i}m}\tilde C^{\underline{j}]n} D_k B_{mn}-2D^{[\underline{i}}B_{km}\tilde
C^{\underline{j}]m}
\\[0.1cm]
R^{ijk}=&-3\tilde D^{[\underline{i}} \beta^{\underline{jk}]}
+3\tilde
D^{[\underline{i}}\tilde A^{\underline{j}\gamma}\tilde
A^{\underline{k}]}{}_\gamma
+3\tilde C^{[\underline{i}m}\tilde D^{\underline{j}} B_{mn}\tilde
C^{\underline{k}]n}
}
For $A_i{}^\alpha=\tilde A^i{}_\alpha=0$, these expressions are consistent with
the ones derived in \cite{Aldazabal:2011nj,Geissbuhler:2013uka, Blumenhagen:2013hva}.

Due to  the extra gauge coordinates $y^\alpha$ in heterotic DFT,
there exist new types of fluxes.
Choosing at least one index
of ${\cal F}_{ABC}$ to be a gauge index,  the antisymmetry $\Omega_{ABC}=-\Omega_{ACB}$
in all indices forces us to set either $\beta^{ij}=\tilde
A^i{}_{\alpha}=0$ or $B_{ij}=A_i{}^{\alpha}=0$. Of course, one can
choose these constraints independently for each direction $(ij)$ or $(i)$,
respectively. In the following, we present the result for choosing the
same set of conditions for all directions.

Thus,  in the geometric frame $\beta^{ij}=\tilde A^i{}_{\alpha}=0$,
 we get the following three   types of non-vanishing gauge fluxes
\eq{
G_{\alpha ij}
&=-2D_{[\underline{i}}A_{\underline{j}] \alpha}-D_\alpha B_{ij} +D_\alpha A_{[\underline{i}}{}^\gamma
A_{\underline{j}]\gamma}\\
J^j{}_{\alpha i}&=\tilde\partial^j A_{i\alpha}\\
K_{\alpha\beta i}&=2\,D_{[\underline{\alpha}}
A_{i\underline{\beta}]}\, .
}
Solving the strong constraint via
$\tilde\partial^i=\partial_\alpha=0$, the first flux
reduces to the familiar form of the field strength
\eqref{fieldstrengthG}
for an  abelian field.
In the non-geometric frame $B_{ij}= A_i{}^{\alpha}=0$, the
non-vanishing  fluxes  are
\eq{
J^j{}_{\alpha i}&= -\partial_i\tilde A^j{}_\alpha \\
\tilde G_\alpha{}^{ij}
&=-2\tilde D^{[\underline{i}}\tilde
A^{\underline{j}]}{}_\alpha
-D_\alpha \beta^{ij}+D_\alpha \tilde A^{[\underline{i}\gamma}\tilde
A^{\underline{j}]}{}_{\gamma} \\
\tilde K^{\alpha\beta i}&=2\,D^{[\underline{\alpha}}
\tilde A^{i\underline{\beta}]}\, .
}
Hence, the flux $J^j{}_{\alpha i}$ in the non-geometric frame
is indeed the flux we encountered in the previous section after
 applying one T-duality.     Similarly, reducing  $\tilde
 G_\alpha{}^{ij}$  for $\partial_i=\partial_\alpha=0$, one obtains
\eq{
\tilde G_\alpha{}^{ij}=-2\tilde\partial^{[\underline{i}}\tilde
A^{\underline{j}]}{}_\alpha\, ,
}
the gauge flux of $\tilde A$ found in the background after applying
two T-dualities.

For a non-holonomic basis, one finds the commutators
\eq{&[\partial_a, \partial_b]=f^c{}_{ab}\, \partial_c\,,\quad{\rm
    with}\quad  f^c{}_{ab}:= e_i{}^c
(\partial_a e_b{}^i- \partial_b e_a{}^i).\\
&[\tilde \partial^a, \tilde\partial^b]=\tilde f_c{}^{ab}\,
\tilde\partial^c\, ,\quad{\rm with}\quad
\tilde f_c{}^{ab}:= e_c{}^i (\tilde\partial^a e_i{}^b-
\tilde\partial^b e_i{}^a)
}
providing correction terms to the fluxes  shown above.
The resulting  rather lengthy expressions for these fluxes  can be found in appendix \ref{app_fluxhet}.

The upshot of the explicit analysis of this section is that, for
the heterotic string,  the
T-dual of the constant gauge flux background on a flat geometry
is a {\it non-geometric} background.
Therefore, the concept of
non-geometry does not only apply to closed string three-form
backgrounds but also to gauge flux backgrounds.
Moreover, we have seen that for the description of these T-dual
backgrounds, it is appropriate to change to a non-geometric
frame, where in particular the gauge 1-form $A=A_i\,dx^i$ is replaced
by a gauge 1-vector $\tilde A=\tilde A^i\, \partial_i$.

\subsection{Comment on $R$-flux and non-associativity}

It has been suggested that the non-geometric $R$-flux background gives
rise to some non-associativity of the usual coordinates \cite{Bouwknegt:2004ap,Blumenhagen:2010hj,Lust:2010iy,Blumenhagen:2011ph,Mylonas:2012pg}. In the
context of DFT this was analyzed in \cite{Blumenhagen:2013zpa}, where it was studied how
DFT behaves if  one defines a new tri-product given in terms of the
three-index flux as
\eq{
  f\,\triangle\, g\,\triangle\, h= f\, g\, h\, + {\cal F}_{ABC}\, D^A f\,
   D^B g\, D^C h+\ldots \, .
}
For objects satisfying the strong constraint, the correction
identically vanishes so that, in particular, for the action such a
deformation has no effect (see \cite{Blumenhagen:2013zpa} for more details).

Since the coordinates themselves are not conformal fields, one does not
necessarily expect them to satisfy the level-matching constraint and
consequently not the strong constraint.
The implied tri-bracket $[x^i,x^j,x^k]$ for the coordinates is
governed by the non-geometric flux coupled to just ordinary
derivatives.  Thus, we  are focusing on the term
\eq{
\label{ncaa}
   {\cal F}_{ABC}\, D^A f\,
   D^B g\, D^C h = \rho^{ijk}\, \partial_i f\, \partial_j
   g\, \partial_k h+\ldots
}
which for usual DFT was just
$\rho_{bos}^{ijk}=3\tilde \partial^{[\underline{i}}
\beta^{\underline{jk}]}$.
The natural expectation is that, in the heterotic case, this gets
generalized to the gauge invariant combination\footnote{We confirmed this behavior by performing a conformal field theory
analysis along the lines of \cite{Blumenhagen:2011ph}.}
\eq{
   \rho_H^{ijk}=3\Big(\tilde \partial^{[\underline{i}} \beta^{\underline{jk}]}
-\tilde
\partial^{[\underline{i}}\tilde A^{\underline{j}\gamma}\tilde
A^{\underline{k}]}{}_\gamma\Big)\, .
}
However, evaluating \eqref{ncaa} for  a  holonomic non-geometric frame, one finds
\eq{
\label{nonasso}
\rho^{ijk}= 3\big(\tilde\partial^{[\underline{i}} E_{A}{}^{\underline
    j}\big)  E^{A\underline{k}]} = -3\Big(\tilde \partial^{[\underline{i}} \beta^{\underline{jk}]}
+\tilde
\partial^{[\underline{i}}\tilde A^{\underline{j}\gamma}\tilde
A^{\underline{k}]}{}_\gamma\Big)\, ,
}
showing that the relative sign
between the two terms on the right hand side of \eqref{nonasso} is
different.
As a consequence, this object $\rho^{ijk}$
is not invariant under ${\tilde A}$ gauge transformations
$\tilde A^i{}_\alpha=\tilde A^i{}_\alpha+\tilde\partial^i \lambda_\alpha$, unless
the non-geometric gauge flux $\tilde G_\alpha{}^{ij}=
\tilde  \partial^{[\underline{i}}\tilde A^{\underline{j]}}{}_\alpha$
vanishes. We observe that this sign flip can be reconciled with
heterotic DFT by defining instead
\eq{
  f\,\triangle\, g\,\triangle\, h= f\, g\, h\, + {\cal F}^H_{ABC}\, D^A f\,
   D^B g\, D^C h+\ldots \,
}
with
\eq{
{\cal F}^H_{ABC}(\beta,\tilde A)={\cal F}_{ABC}(\beta,\tilde A)-2\, {\cal
  F}_{ABC}(\beta,\tilde A=0)\, .
}

\subsection{Comment on S-duality}

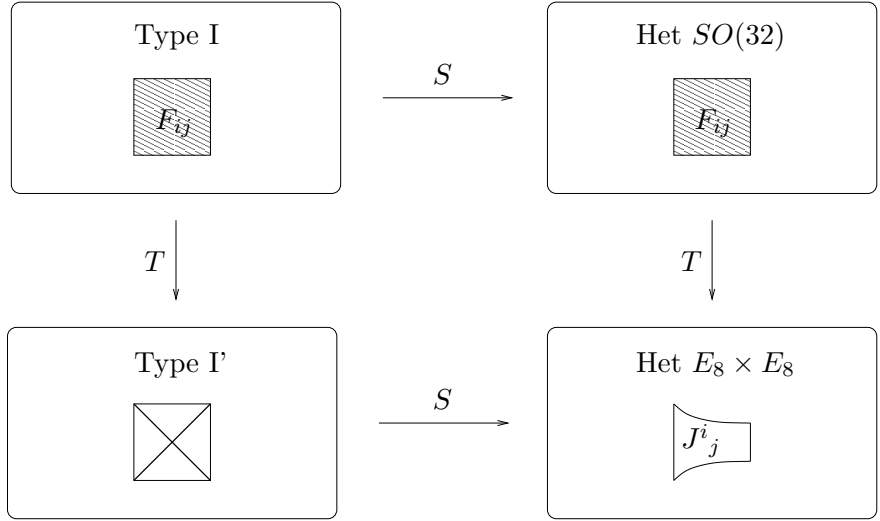
\begin{figure}[ht]
  \centering
  \includegraphics[width=0.8\textwidth]{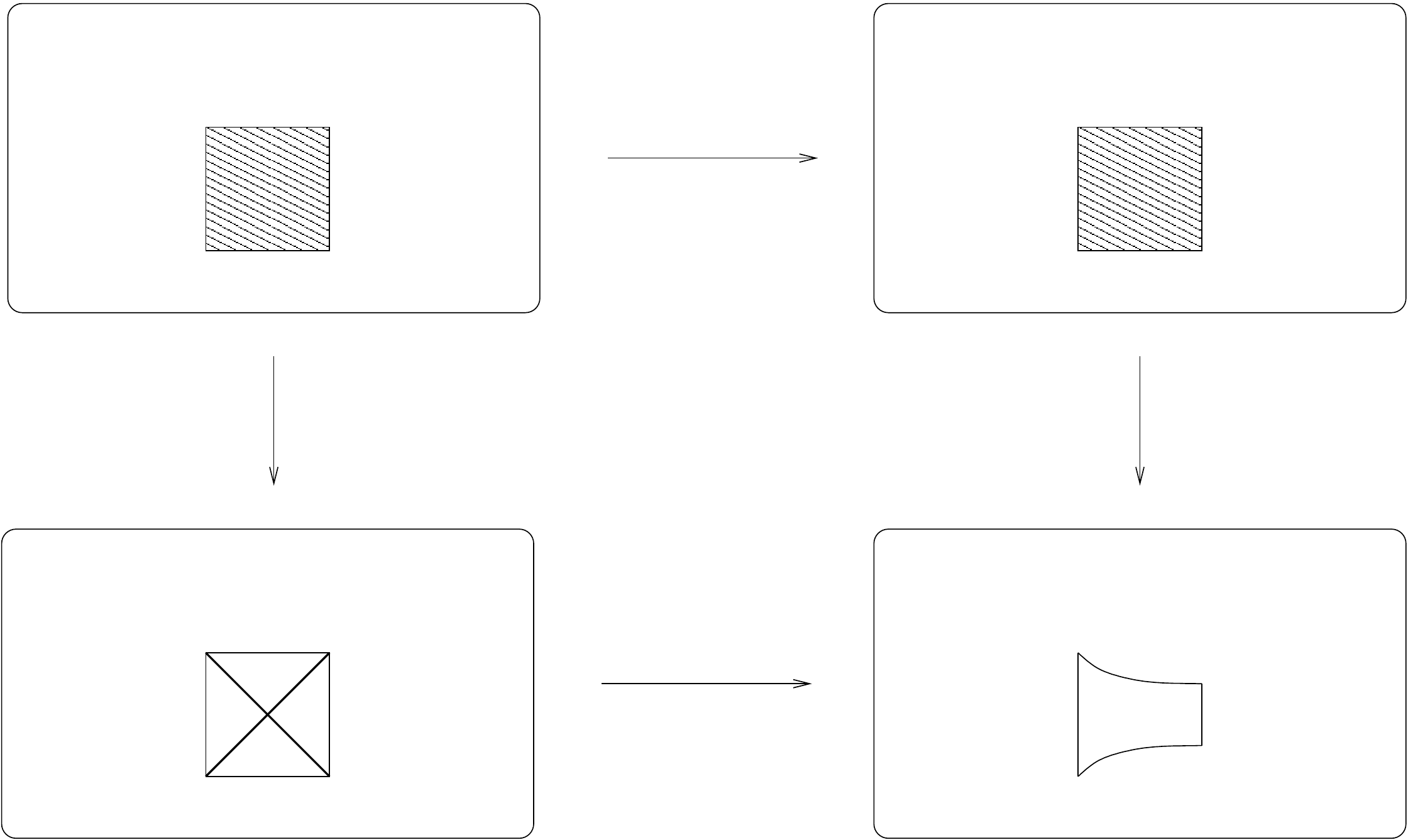}
  \begin{picture}(0,0)
  \put(-285,180){Type I}
  \put(-95,180){Het $SO(32)$}
  \put(-285,56){Type I'}
  \put(-95,56){Het $E_8\times E_8$}
  \put(-277,148){$F_{ij}$}
  \put(-73,148){$F_{ij}$}
   \put(-78,27){$J^i{}_j$}
   \put(-172,165){$S$}
   \put(-172,42){$S$}
   \put(-281,94){$T$}
   \put(-78,94){$T$}
  \end{picture}
  \caption{\small $S$- and $T$-duality between Type I and heterotic string.}
  \label{sdual}
\end{figure}

Let us now consider the $SO(32)$ heterotic string compactified on a
two-torus with constant abelian gauge flux $F=F_{12}$. This configuration
is known to be S-dual to the Type I string \cite{Polchinski:1995df}
compactified on a two-torus
where the $D9$-brane carries the same gauge flux $F$. Applying a
T-duality in the $y$-direction to this latter configuration yields
the Type I' string with a $D8$-brane at an angle with respect to the
$O8$-planes. One might ask whether there exist an S-dual to this
configuration.  The answer to this question is not obvious, as
in the heterotic string there are no $8$-branes.
However, recall that we have just seen that the
T-dual to the $SO(32)$ heterotic string with gauge flux is
a non-geometric background of the $E_8\times E_8$ heterotic string
carrying flux $J=J^1{}_2$. Therefore, by completing the diagram as
shown in figure \ref{sdual} we are led to the conjecture that
the S-dual of the $D8$-brane at angle in Type I' is a non-geometric $J$-flux
background of the heterotic string.

\section{A Lie algebroid for heterotic field redefinitions}
\label{sec:redefinition}

In the previous section we have seen that a field redefinition can
help in simplifying the description of non-geometric backgrounds.
As we discussed in the introduction, this fact is familiar from $O(D,D)$ generalized geometry and DFT,
respectively.
Recall that in \cite{Blumenhagen:2013aia} the general structure of $O(D,D)$ induced field definitions
was clarified in the framework of generalized geometry. The two main
results were that for every such field redefinition, one can associate a
corresponding Lie algebroid so that the redefined supergravity action
is governed by the differential geometry of that Lie algebroid.

In this section, we show that this picture also holds for the
heterotic case, i.e. to every $O(D,D+n)$ induced field redefinition one
can associate a corresponding Lie algebroid so that in the new field
variables the heterotic action  is governed by the
differential geometry of that Lie algebroid. For the definition
of a Lie algebroid, please consult appendix \ref{app_lie}.
We will also show that the non-geometric frame \eqref{genmetricnongeo} does
also fit into this scheme.
Since the story is very similar, we will be rather
brief here and refer the reader to \cite{Blumenhagen:2013aia} for more information on
Lie algebroids and its differential geometry.

\subsection{$O(D,D+n)$-induced field redefinition}

In abelian heterotic generalized geometry,  one  considers a D-dimensional
manifold $M$ with usual coordinates $x^i$,  equipped with a generalized bundle $E=TM\oplus T^*M\oplus V$, whose
sections are formal sums  $\xi+\tilde \xi+ \lambda$ of vectors,
$\xi=\xi^i (x) \,\partial_i$,
one-forms, $\tilde \xi=\tilde \xi_i(x)\, dx^i$  and gauge transformations,
$\lambda=(\lambda_1(x),\ldots \lambda_n(x))$,   of $U(1)^n$.
On this bundle one defines a generalized ${\cal H}_{MN}$ metric taking the familiar
form \eqref{genmetric} in terms of the fundamental fields
$g_{ij}$, $B_{ij}$ and $A_i{}^\alpha$.
An  $O(D,D+n)$ transformation ${\cal M}$ acts on the generalized
metric via conjugation, i.e.
\eq{
\label{genmetricfieldred}
{\hat H}(\hat g, \hat B, \hat A)={\cal M}^t\, H(g,B,A)\,{\cal M}
}
and therefore defines a field redefinition
\eq{
   (g,B,A)\longrightarrow (\hat g, \hat B, \hat A)\, .
}
The heterotic action in terms of the fields $(g,B,A)$ is the heterotic
supergravity action \eqref{effaction}. The question is how the action in the
new field variables $(\hat g, \hat B, \hat A)$ looks.  Just inserting
the field redefinition gives a plethora of terms so that an
organizing principle is needed.

To proceed, we write  a general $O(D,D+n)$ matrix  ${\cal M}$ as
\eq{
    {\cal M}=
      \left(\begin{matrix}
      a& b&  m\\
      c&d&n \\
      p&q&z
\end{matrix}\right)\, .
}
This transformation has to leave the $\eta$ metric \eqref{eta}
invariant, i.e.
\eq{
{\cal M}^t\, \eta\,{\cal M}=\eta\, ,
}
leading to  six independent constraints on the submatrices
\eq{
\label{transmatrix}
c^t a +a^t c +p^t p=&0\\
c^t b +a^t d +p^t q=&1\\
c^t m +a^t n +p^t z=&0\\
d^t b +b^t d +q^t q=&0\\
d^t m +b^t n +q^t z=&0\\
n^t m +m^t n +z^t z=&1\, .
}
Now applying \eqref{genmetricfieldred} we can read off the induced
field redefinition. For the upper-left component of
${\hat H}(\hat g, \hat B, \hat A)$  one obtains
\eq{
{\hat H}(\hat g, \hat B, \hat A)_{ul}
=\Big[a-A\, p-\big(g+B+{\textstyle \frac{1}{2}} A^2\big)\, c\Big]^t\, g^{-1}\,\Big[a-A\,
p-\big(g+B+{\textstyle \frac{1}{2}} A^2\big)\, c\Big]
}
which, comparing with general form of the  generalized metric, gives
$\hat g^{-1}$. Thus, we get
\eq{
\label{anchor_g}
\hat g=(\gamma^{-1})\, g \,(\gamma^{-1})^t
}
where the matrix $\gamma$ is given as
\eq{
\label{anchor_gamma}
\gamma=a-A\, p-\big(g+B+{\textstyle\frac{1}{2}} A^2\big) \,c\, .
}
In order to consider the redefined Kalb-Ramond field $\hat B$ which is
contained in $\hat C$, we consider  the upper-middle component of the
redefined generalized metric
\eq{
{\hat H}(\hat g, \hat B, \hat A)_{um}
=1+\Big[a-A\, p-\big(g+B+{\textstyle \frac{1}{2}} A^2\big)\,
c\Big]^t\, g^{-1}\,
\Big[b-A\, q-\big(g+B+{\textstyle \frac{1}{2}} A^2\big)\, d\Big]
}
and compare it with $\hat {H}_{um}=-\hat g^{-1} \hat C$.
Thus, we find
\eq{
\label{anchor_c}
\hat C=(\gamma^{-1})\, \frak{C}
\,(\gamma^{-1})^t\, \qquad {\rm with}\quad  \frak{C}=\delta\, \gamma^t-g
}
with the matrix $\delta$ defined as
\eq{
\label{anchor_delta}
\delta=-b+A\, q+\big(g+B+{\textstyle \frac{1}{2}} A^2\big)\, d\, .
}
It remains to determine the $O(D,D+n)$ induced field redefinition for
the gauge field $A$.
For that purpose,  we look into the upper-right element of
 the generalized metric
\eq{
{\hat H}(\hat g, \hat B, \hat A)_{ur}
=\Big[ a-A\, p-\big( g+B+{\textstyle \frac{1}{2}} A^2\big)\, c\Big]^t
\, g^{-1}\, \Big[ m-A\,  z-\big(g+B+{\textstyle \frac{1}{2}} A^2\big)
\, n\Big]
}
and identify it with $-\hat g^{-1} \hat A$.  Thus,  we obtain
\eq{
\label{anchor_a}
\hat A=(\gamma^{-1})\, \frak A
}
with
\eq{\label{anchor_zeta}
\frak A=-m+A\, z+\big(g+B+{\textstyle \frac{1}{2}}\, A\big)\, n\, .
}
From $\frak{C}$ and $\frak{A}$ one can define also a new $B$-field $\frak{B}$ via
\eq{
\label{anchor_b}
\hat B=(\gamma^{-1})\, \frak{B}
\,(\gamma^{-1})^t\, \qquad {\rm with}\quad  \frak{B}=\frak{C}-{1\over 2}\frak{A}\otimes \frak{A}\,.
}

Thus the field redefinition is of a very peculiar form, where the
matrix $\gamma$ plays a prominent role. In fact, the structure of
the field redefinition of $g$ and $B$ is completely analogous to
\cite{Blumenhagen:2013aia},
only containing some new gauge field dependent corrections in $\gamma$
and $\delta$. Thus it is straightforward to proceed as in \cite{Blumenhagen:2013aia}
and to identify
\eq{
\rho=(\gamma^{-1})^t
}
as the anchor map of a Lie algebroid (see appendix \ref{app_lie}).

This Lie algebroid lives on the
tangent bundle itself, i.e. $E=TM$ and the anchor map $\rho: E\to TM$
 acts on  a vector field $X=X^i \partial_i\in E$ as\footnote{Here we present the relations in a holonomic basis. For
the non-holonomic case, we refer to \cite{Blumenhagen:2013aia}.}
\eq{
  \rho (X)= (\rho^i{}_j\, X^j)\, \partial_i = X^i (\rho^t)_i{}^j \, \partial_j = X^i\, D_i \,,
}
where we defined the partial derivative for the Lie algebroid
as
\eq{	
\label{new_partial}
   D_i = (\rho^t)_i{}^j\,  \partial\vphantom{(\rho^t)}_j \,.
}
The bracket $\lb \cdot,\cdot \rb$ on $E=TM$ is defined as
\eq{\label{tbracket}
  \lb X,Y\rb = \Big(X^j D_j Y^k-Y^j D_j X^k +X^i\,Y^j\,F_{ij}{}^k\Big)\, \partial_k \,.
}
with the structure constants
\eq{
  \label{struct_const}
  F_{ij}{}^k =
  (\rho^{-1})^k{}_m
  \Bigl( D_i (\rho^t)_j{}^m - D_j (\rho^t)_i{}^m \Bigr)
  \,.
}
Indeed, this bracket satisfies the homomorphism property
\eq{
  \rho\big( \lb  X,Y\rb\big)=[\rho (X),\rho (Y)] \,.
}
Furthermore, by construction the new bracket $\lb\cdot,\cdot\rb$ satisfies the
Jacobi identity \eqref{jacobii} as well as the
Leibniz rule \eqref{Leibniz}.
Thus, for every $O(D,D+n)$ induced field redefinition we  have
associated a corresponding  Lie algebroid. The true power of this formal
approach will become clear in the next section.

\subsection{The redefined heterotic action}
Recall that the NS-sector of the heterotic DFT action is
\eq{\label{effaction1}
{\cal S}=\int dx \sqrt{g} e^{-2\phi}\Big(R+4(\partial
\phi)^2-\frac{1}{12}H^{ijk}{H}_{ijk}-\frac{1}{4}{G}^{ij\alpha}{G}_{
ij\alpha}\Big)
}
with  the
three-form $H=dB-{1\over 2} \delta_{\alpha\beta} A^\alpha\wedge
dA^\beta$ and the abelian two-form field strength $G^\alpha=dA^{\alpha}$.
As  derived in detail in \cite{Blumenhagen:2013aia},  the field redefinition
is completely given by pulling indices up and down by the action of
the anchor (here $\rho^t=\gamma^{-1}$).
For the metric we found \eqref{anchor_g}, which implies that the
quantities in the gravitational sector transform as
\eq{\label{redmitanchor}
\hat R^q{}_{mnp}&= (\rho^{-1})^q{}_l \,\rho^{i}{}_m \,\rho^{j}{}_n
\,\rho^{k}{}_p\, R^l{}_{ijk}\, ,\qquad
 \hat R_{mn}= \rho^i{}_m\, \rho^j{}_n \, R_{ij}\, ,\\[0.1cm]
\hat R&= R\, ,\qquad \sqrt{|\hat g|}=\sqrt{|g|}|\rho^t|\, ,\qquad  \hat \phi= \phi
}
where the derivative for the transformed theory is
\eqref{new_partial}.

For the flux sector, so far we know the transformation behavior of
gauge potentials $B$ and $A$. Therefore, one still needs to find the
proper definition of the new field strengths so that they also
transform properly, i.e. just by  pulling up and down indices with the
anchor. For that purpose one needs to invoke the Lie algebroid
differential $d_E$ defined in appendix \ref{app_lie}.
For the gauge field strength $G=dA$, using  the
relation \eqref{diffs} one can show
\eq{
(\Lambda^2 \rho^*) d_E \hat{\frak{A}} = d( \rho^* \hat{\frak{A}})=dA
}
with $\rho^*=(\rho^t)^{-1}=\gamma$, so that
\eq{
         \hat{\frak{G}}:= d_E \hat{\frak{A}}=(\Lambda^2 \rho^t)\, G
}
is the correct definition of the transformed field strength that
transforms
properly.
Analogously, one can show
\eq{
         d_E \hat{\frak{B}}=(\Lambda^3 \rho^t)\, dB
}
so that the proper  three-form flux is given by
\eq{
         \hat{\frak{H}}:= d_E \hat{\frak{B}}-{1\over 2} \hat{\frak{A}}\wedge
         d_E \hat{\frak{A}}=(\Lambda^3 \rho^t)\, H\, .
}
Its  Bianchi identity reads
\eq{
     d_E \hat{\frak{H}}=-{1\over 2} \hat{\frak{G}}\wedge
     \hat{\frak{G}}\, .
}
Thus, each quantity appearing  in the heterotic action \eqref{effaction1} now
transforms properly so that  the action in the redefined fields can be
expressed as
\eq{\label{redaction}
{\cal S}=\int dx \sqrt{\hat g}\,\,|\rho^*|\,\, e^{-2\phi}\Big(\hat
R+4(D
\phi)^2-\frac{1}{12} \hat{\frak{H}}^{ijk}\hat{\frak{H}}_{ijk}-\frac{1}{4} \hat{\frak{G}}^{ij\alpha}
\hat{\frak{G}}_ {ij\alpha}\Big)\, .
}
This  has the analogous form as the original action, but with the new
fields defined in the framework of  the differential geometry of the Lie algebroid.
Therefore, the latter provides the organizing  principle for expressing
the action in $O(D,D+n)$ induced redefined field variables.

Note that the symmetries of this action are just the transformed
diffeomorphisms and $B$- and $A$-field gauge transformations of the original
action. Clearly, just by a field redefinition, one does not gain new
symmetries.
Therefore, the $\tilde A$ field gauge transformation \eqref{concretegaugetrafong}, needed for the
transition function of the T-dual non-geometric $J$-flux background is
{\it not} a symmetry of \eqref{redaction}.  Thus, as in
generalized geometry \cite{Blumenhagen:2013aia}, a field redefinition helps to bring in each
patch a non-geometric background in a simple form, but in general it does not provide
a global description of the background.

\subsection{The non-geometric frame}
In this section we show that the field redefinition between the
geometric and the non-geometric frame from section \ref{sec_tduali} can also
be described in this framework.
For that purpose, first recall the form of the generalized metric in
these
two frames. In the geometric one, we have
\eq{
{\cal H}_{MN} =
 \begin{pmatrix}
  g^{ij} & -g^{ik}C_{kj} & -g^{ik}A_{k\beta} \\[0.1cm]
  -g^{jk}C_{ki} & g_{ij}+C_{ki}g^{kl}C_{lj}+A_i{}^\gamma A_{j\gamma} &
C_{ki}g^{kl}A_{l\beta}+A_{i\beta} \\[0.1cm]
  -g^{jk}A_{k\alpha }  & C_{kj}g^{kl}A_{l\alpha}+A_{j\alpha}  &
\delta_{\alpha\beta}+A_{k\alpha}g^{kl}A_{l\beta}
 \end{pmatrix}
}
and in the non-geometric one
\eq{
{\cal H}_{MN} =
 \begin{pmatrix}
\tilde g^{ij}+\tilde C^{ki}\, \tilde g_{kl}\,\tilde C^{lj}+\tilde A^i{}_\gamma
\,\tilde A^{j\gamma} & -\tilde g_{jk}\,\tilde C^{ki}
&\tilde C^{ki}\, \tilde g_{kl}\,\tilde A^{l}{}_{\beta}+\tilde
A^i{}_{\beta} \\[0.1cm]
  -\tilde g_{ik}\,\tilde C^{kj} &  \tilde g_{ij} & -\tilde g_{ik}\,\tilde
A^k{}_{\beta} \\[0.1cm]
  \tilde C^{kj}\, \tilde g_{kl}\,\tilde A^{l}{}_{\alpha}+\tilde A^j{}_{\alpha}  &
-\tilde g_{jk}\,\tilde A^k{}_{\alpha } &
\delta_{\alpha\beta}+\tilde A^k{}_{\alpha}\, \tilde g_{kl}\,\tilde
A^l{}_{\beta}
 \end{pmatrix}\, .
}
By comparison of the components, the corresponding field redefinition takes the form
\eq{\label{nongeomfieldref}
\tilde g&=g+C^t\, g^{-1} C+A^2\\
\tilde C&=\tilde g^{-1}\,  C^t\, g^{-1}\\
\tilde A&=
       -(\tilde g^{-1}+\tilde C)\, A\, .
}
Analogous to \cite{Blumenhagen:2013aia}, we propose that this transformation is implemented by choosing
\eq{{\cal M}=
 \begin{pmatrix}
0 & \tilde g & 0\\
\tilde g^{-1} & 0 & 0\\
0 & 0 & 1
 \end{pmatrix}
}
with $\tilde g=g+C^t\, g^{-1} C+A^2$.
Evaluating the expressions
\eqref{anchor_gamma},\eqref{anchor_c},\eqref{anchor_delta},
\eqref{anchor_zeta} we obtain as intermediate results
\eq{\label{anchors}
\gamma&=-(g+C)\,\tilde g^{-1}\, \quad{\rm so\ that}\quad \gamma^{-1}=-(g+C^t) g^{-1} \,,\\
\delta&=-\tilde g\, \qquad\phantom{,aaaaae} {\rm so\ that}\quad \frak{C}=C^t \,,\\
\frak{A}&=A\, .
}
Using these relations further in \eqref{anchor_g}, \eqref{anchor_c} and \eqref{anchor_a}
we finally get
\eq{
\label{anchor formula}
\hat g&=(\gamma^{-1})\, g \,(\gamma^{-1})^t =\tilde g\\
\hat C&=(\gamma^{-1})\, \frak{C}\,(\gamma^{-1})^t
=C^t\, g^{-1}\, \tilde g\\
\hat A&=(\gamma^{-1})\, \frak{A} =-(1+C^t\, g^{-1})A\, .
}
Here $\hat C$ and $\hat A$ are still forms. For transforming  them
into a bi-vector and a vector, one pulls up the indices with $\tilde g^{-1}$ so that
\eq{
\tilde C&=\tilde g^{-1}\,\hat C \,\tilde g^{-1}=\tilde g^{-1}\,C^t\, g^{-1}\\
\tilde A&=\tilde g^{-1}\,\hat A=-(\tilde g^{-1}+\tilde C)\,A\, ,
}
which precisely agrees with the field redefinition of the non-geometric
frame \eqref{nongeomfieldref}.


\section{Conclusion}

In this paper we studied a couple of aspects of heterotic DFT in more
detail. We think that, while the general formalism of heterotic DFT was developed
before and is a  straightforward generalization of bosonic DFT, the
concrete evaluation of its consequences, in particular for issues
related to the gauge field, deserved a further study.

Indeed, by applying the T-duality rules ($\alpha'$ corrected heterotic
Buscher rules)  to a flat
background with a constant gauge field, we found  non-geometric
backgrounds, which were very similar to the $Q$- and $R$-flux
backgrounds in bosonic DFT.
Namely, after one T-duality we already obtained a background which was
best described by changing to a non-geometric frame, where the gauge
one-form has turned into a gauge one-vector. The required transition function
between two patches was given by a new symmetry, namely a one-vector gauge
transformation involving a winding dependence.
Thus, this background is globally non-geometric, an effect introduced
by the $\alpha'$ corrected Buscher rules. Applying a further
T-duality, the arising  background was even locally non-geometric.

Even though, we were only considering abelian gauge fields, we expect
this picture to generalize also to non-abelian gauge fields. The
latter
are introduced via a gauging procedure that generically breaks the $O(D,D+n)$
symmetry to $O(D,D)$. However, T-duality is a special element of
$O(D,D)$ so that it can still be treated analogously to the abelian case.

Moreover, we clarified which type of fluxes are turned on in these
backgrounds and how they are microscopically described in terms
of the fundamental fields in the theory. We argued that the constant
non-geometric $J$-flux background of the $E_8\times E_8$ heterotic
string can be considered the S-dual  of a Type I' background with
a $D8$-brane intersecting the $O8$-plane at an angle.

Led by the apparent necessity of field redefinitions, we considered the general
question what effect an $O(D,D+n)$ induced  field redefinition has on the heterotic
supergravity action. Generalizing \cite{Blumenhagen:2013aia}, we investigated this
question in the framework of generalized geometry and found very
similar results, though now  including various  corrections due to the
present one-form gauge field. In particular, the organizing  principle
for the terms in the redefined action was given by the differential geometry of a Lie algebroid, whose
anchor was related to the $O(D,D+n)$ transformation.
The non-geometric frame was identified with just a specific $O(D,D+n)$
induced field redefinition.

Even though, here we were only considering the NS part of the heterotic
action, we expect that the whole action including the fermionic terms
are governed by the objects in the differential geometry of the Lie
algebroid. This includes e.g. the kinetic terms for the gravitinos and
gluinos, that involve a spin-connection. Moreover, here we were neglecting the gravitational
Chern-Simons term (see \cite{Bedoya:2014pma} for a recent treatment in
DFT).  Introducing non-abelian gauge fields via gauging, breaks the
$O(D,D+n)$
symmetry so that in this case only the remaining symmetry should
be used for a field redefinition.

\vspace{0.3cm}

\noindent
\section*{Acknowledgments}
We would like to thank Andre Betz, Michael Fuchs,
Daniela Herschmann, Dieter L{\"u}st, Felix Rennecke and
Christian Schmid for discussions.
R.B. would like to express a special thanks to
the Mainz Institute for Theoretical Physics (MITP)
for its hospitality and support.
R.S. would also like to thank
Klaus Altmann for support. The work of R.S. is supported by  the
China Scholarship Council (CSC).

\vspace{0.5cm}

\appendix

\section{The Buscher rules derived from heterotic DFT}
\label{app_b}

Using the implementation of T-duality in heterotic DFT, one can now
quite generally (re-)derive the Buscher from the conjugation of
the generalized metric with the corresponding T-duality matrix.
Carrying out this procedure for a T-duality in the $x^\theta$
direction, we get precisely
the $\alpha'$ corrected Buscher rules presented
in \cite{Serone:2005ge}
\eq{
\label{Buscher}
G_{\theta\theta}^\prime&=\frac{G_{\theta\theta}}{\left(G_{\theta\theta}+{\alpha'\over
      2} A_\theta^2\right)^2}\\[0.2cm]
G_{\theta i}^\prime&=-\frac{G_{\theta\theta} B_{\theta i}+{\alpha'\over 2} G_{\theta i}
  A_\theta^2-{\alpha'\over 2} G_{\theta\theta}\,A_\theta
  A_i}{\left(G_{\theta\theta}+{\alpha'\over 2} A_\theta^2 \right)^2}\\[0.2cm]
G_{ij}^\prime&=G_{ij}-\frac{G_{\theta i}G_{\theta j}-B_{\theta
    i}B_{\theta j}}{\left(G_{\theta\theta}+{\alpha'\over
      2} A_\theta^2\right)}\\[0.1cm]
&\quad  -\frac{1}{\left(G_{\theta\theta}+{\alpha'\over 2} A_\theta^2 \right)^2}
\bigg( G_{\theta\theta}\Big[ {\textstyle {\alpha'\over 2}} B_{\theta j}  A_\theta
  A_i+{\textstyle {\alpha'\over 2}} B_{\theta i} A_\theta A_j-{\textstyle {{\alpha'}^2\over 4}} A_\theta A_i\;
  A_\theta A_j\Big]\\[0.1cm]
&\qquad\quad +{\textstyle {\alpha'\over 2}} A_\theta^2  \Big[(G_{\theta i}-B_{\theta i}) (G_{\theta
    j}-B_{\theta j})+{\textstyle {\alpha'\over 2}} (G_{\theta i} A_\theta A_j +G_{\theta
    j} A_\theta  A_i ) \Big]\bigg)\\[0.3cm]
B_{\theta i}^\prime&=-\frac{G_{\theta i}+{\alpha'\over 2} A_\theta A_i}{\left(G_{\theta\theta}+{\alpha'\over
      2} A_\theta^2\right)}\\[0.2cm]
B_{ij}^\prime&=B_{ij}-\frac{(G_{\theta i}+{\alpha'\over 2}A_\theta
  A_i)B_{\theta j}-(G_{\theta j}+{\alpha'\over 2} A_\theta A_j) B_{\theta i}}{\left(G_{\theta\theta}+{\alpha'\over
      2} A_\theta^2\right)}\\[0.3cm]
{A'}_{\theta}{}^{\alpha}&=-\frac{A_\theta{}^{\alpha}}{\left(G_{\theta\theta}+{\alpha'\over
      2} A_\theta^2\right)}\\[0.2cm]
{A'}_i{}^\alpha&=A_i{}^{\alpha}-A_\theta{}^{\alpha}\frac{G_{\theta
    i}-B_{\theta i}+{\alpha'\over 2} A_\theta A_i}{\left(G_{\theta\theta}+{\alpha'\over
      2} A_\theta^2\right)}
}
where e.g. $A_\theta A_i=A_\theta^\alpha  A_{i\alpha}$. Here the
metric and the Kalb-Ramond field have  dimension $[l]^0$  and the gauge
field $[l]^{-1}$.

\section{Non-holonomic fluxes for heterotic DFT}
\label{app_fluxhet}

In this appendix we present the explicit expressions of the fluxes in a a
non-holonomic basis. From the generalized vielbein $E_A{}^M$ and the dilation
$d$ one can build the generalized fluxes
\eq{\label{genflux}
{\cal F}_{ABC}&=E_{CM} {\cal L}_{E_A}
E_B{}^M=\Omega_{ABC}+\Omega_{CAB}-\Omega_{BAC}\\
{\cal F}_{A}&=-e^{2d}{\cal L}_{E_A} e^{-2d}=-\partial_{M}E_A{}^M+2D_A
d\, .
}
The generalized derivative
$D_{A}=E_{A}{}^{M} D_M $ takes the form
\eq{
\tilde D^a &= \tilde \partial^a +\tilde C^{am}C_{mn} \tilde \partial^n
          -\tilde C^{am} \partial_m -\tilde A^{a\gamma}\partial_{\gamma}\cr
 D_a &=\partial_a - C_{am}\tilde\partial^m-A_a{}^\gamma\partial_{\gamma}\cr
D_{\alpha} &=\partial_{\alpha}+A_{m\alpha}\tilde\partial^{m}+\tilde
          A^m{}_{\alpha}\partial_m\, .}
As in section $3$, we present the geometric fluxes for the physically relevant
case of $\tilde A^a{}_\alpha=0$ and the non-geometric fluxes for
$A_a{}^\alpha=0$.
From the flux definition \eqref{genflux} we obtain the geometric fluxes
\footnote{Note that the
derivative $D^i, D_i$ and $D_\alpha$ will also be simplified.}
\eq{
H_{abc}&=-3\big(D_{[\underline{a}} B_{\underline{bc}]} -D_{[\underline{a}}
A_{\underline{b}\gamma} A_{\underline{c}]}{}^\gamma+
f^m{}_{[\underline{ab}}\,C_{\underline{c}]m} -C_{[\underline{a}m}
C_{\underline{b}n}\tilde{f}_{\underline{c}]}{}^{mn}-A_{[\underline{a}}{}^\beta
\partial_\beta e_{\underline{b}}{}^i C_{{\underline{c}]}i}\big)
\\
F^c{}_{ab}
&=f^c{}_{ab}-\tilde D^c B_{ab}+\tilde D^c A_{[\underline{a}\gamma}
A_{\underline{b}]}{}^\gamma+2C_{[\underline{a}m}\tilde
f_{\underline{b}]}{}^{mc}
-2 D_{[\underline{a}}\beta^{cm}A_{\underline{b}]}{}_\gamma A_m{}^\gamma
-2\beta^{cm}D_{[\underline{a}}C_{m\underline{b}]}\\
&\
+3\beta^{cm} \big(f^n { } _{[\underline{ma}}C_{
\underline{b}]n}- C_{[\underline{m}n}C_{\underline{a}p} \tilde
f_{\underline{b}]}{}^{np}\big)
-2\beta^{cm}\big(C_{mi}A_{[\underline{a}}{}^\beta
\partial_\beta
e_{\underline{b}]}{}^i
+C_{{[\underline{a}}i}A_{\underline{b}]}{}^\beta\,\partial_\beta
e_m{}^i\big)\\
&\ -2A_{[\underline{a}}{}^\beta\, \partial_\beta
e_{\underline{b}]}{}^i e^c{}_i
}
and for $A_a{}^\alpha=0$ the non-geometric fluxes read
\eq{Q_c{}^{ab}
&=-D_c \beta^{ab}+D_c\tilde A^{[\underline{a}\gamma}\tilde
A^{\underline{b}]}{}_\gamma
-2\tilde D^{[\underline{a}}B_{cn}\tilde C^{\underline{b}]n}
-\tilde C^{[\underline{a}m}\tilde C^{\underline{b}]n} D_c{}B_{mn}
+\tilde f_c{}^{ab}+2\tilde
C^{[\underline{a}m}f^{\underline{b}]}{}_{mc}\\
&\ +2B_{cm}\tilde
C^{[\underline{a}n}\tilde
f_n{}^{\underline{b}]m}+2\tilde C^{[\underline{a}m}B_{mn}\tilde
f_c{}^{n\underline{b}]}-3\tilde C^{am}\tilde C^{bn} \big(
B_{[\underline{m}p}f^p{}_{\underline{nc}]}
-B_{[\underline{m}p} C_{\underline{n}q}
\tilde f_{\underline{c}]}{}^{pq}\big)\\
&\
+2\big(B_{cm}\tilde C^{[\underline{a}n}\tilde
A^{\underline{b}]\gamma} \partial_\gamma e_n{}^i e^m{}_i
+\tilde A^{[\underline{a}\gamma} \partial_\gamma
e_c{}^i \,e^{\underline{b}]}{}_i
-\tilde C^{[\underline{a}m}B_{mi}
\tilde A^{\underline{b}]\gamma}\partial_\gamma e_c{}^i\big)
\\
R^{abc}&=-3\tilde D^{[\underline{a}} \beta^{\underline{bc}]}+3\tilde
D^{[\underline{a}}\tilde A^{\underline{b}\gamma}\tilde
A^{\underline{c}]}{}_\gamma
+3\tilde C^{[\underline{a}m}\tilde D^{\underline{b}} B_{mn}\tilde
C^{\underline{c}]n}
+6 \tilde
C^{[\underline{a}m}\tilde C^{\underline{b}n}B_{[\underline{m}p}\tilde
f_{\underline{n}]}{}^{p\underline{c}]}
\\
&\ +3\tilde C^{am}\tilde C^{bn}\tilde C^{cp}\big(
B_{[\underline{m}q}f^q{}_{\underline{np}]}-
B_{[\underline{m}q}B_{\underline{n}l} \tilde
f_{\underline{p}]}{}^{ql}\big)
+3\big(
\tilde C^{[\underline{a}m}\tilde C^{\underline{b}n}f^{\underline{c}]}{}_{mn}
-\tilde C^{[\underline{a}m}f_m{}^{\underline{bc}]}\big)
\\
&\ +3\big( \tilde C^{[\underline{a}m}\tilde C^{\underline{b}n}
B_{ni}\tilde A^{\underline{c}]\gamma}\partial_\gamma e_m{}^i
-2 \tilde C^{[\underline{a}m}\tilde A^{\underline{b}\gamma}\partial_\gamma
e_m{}^i e^{\underline{c}]}{}_i\big)\, .
}
For $A_i{}^\alpha=\tilde A^i{}_\alpha=0$, these expressions coincide with
the ones derived in \cite{Geissbuhler:2013uka} and \cite{Blumenhagen:2013hva}.
Similarly, the fluxes ${\cal F}_A$ can be expanded as
\eq{
F_a&=-\partial_m e_a{}^m+\tilde\partial^m C_{am}+\partial_\alpha A_a{}^\alpha+2
D_a d\\
F^a&=\partial_m \tilde C^{am}-\tilde\partial^m e^a{}_m-\tilde\partial^m(\tilde
C^{an}C_{nm})+\partial_\alpha \tilde A^{a\alpha}+2\tilde D^a d\, .
}

Due to the extra gauge coordinates in heterotic DFT, we also
have the gauge fluxes $G_{\alpha ab}, J^c{}_{\alpha b}$
and $\tilde G_\alpha{}^{ab}$. For  $\tilde A^a{}_\alpha=\beta=0$ they become
\eq{G_{\alpha ab}
=&-D_\alpha B_{ab} +D_\alpha A_{[\underline{a}}{}^\gamma
A_{\underline{b}]\gamma}-2D_{[\underline{a}}A_{
\underline{b}]\alpha}+A_{\alpha m}f^m{}_{ab}
+2 C_{[\underline{a}m } A_{ n\alpha} \tilde f_{\underline{b}]}{}^{mn}
\\
&+2\big(
C_{[\underline{a}i}\partial_\alpha e_{\underline{b}]}{}^i-A_{\alpha i}
A_{[\underline{a}}{}^\gamma \partial_\gamma
e_{\underline{b}]}{}^i
\big)
\\
J^c{}_{\alpha b}
=&\tilde \partial^c A_{ b\alpha}
+A_{m\alpha}\tilde f_b{}^{cm}+\partial_\alpha e_b{}^i e^c{}_i
\\
K_{\alpha \beta a}
=&2D_{[\underline{\alpha}} A_{a\underline{\beta}]}
+A_{m\alpha}A_{n\beta}\tilde f_a{}^{mn}
+2 A_{i[\underline{\alpha}} \partial_{\underline{\beta}]} e_a{}^i\, ,
}
while for $A_a{}^\alpha=B=0$ they can be expanded as
\eq{
J^c{}_{\alpha b}
=&-\partial_b\tilde A^c{}_\alpha+\tilde A^m{}_\alpha f^c{}_{mb}+\partial_\alpha
e_b{}^i e^c{}_i
\\
\tilde G_\alpha{}^{ab}
=&-D_\alpha \beta^{ab}+D_\alpha \tilde A^{[\underline{a}\gamma}\tilde
A^{\underline{b}]}{}_{\gamma}-2\tilde D^{[\underline{a}}\tilde
A_\alpha{}^{\underline{b}]}+\tilde A^m{}_\alpha \tilde f_m{}^{ab}+2\tilde
C^{[\underline{a}m}\tilde A^n{}_\alpha
f^{\underline{b}]}{}_{mn}\\
&+2\big(
\tilde C^{[\underline{a}i}\partial_\alpha
e^{\underline{b}]}{}_i-\tilde A^i{}_{\alpha}
\tilde A^{[\underline{a}}{}^\gamma \partial_\gamma
e^{\underline{b}]}{}_i
\big)\\
\tilde K^{\alpha \beta a}
=&2D^{[\underline{\alpha}} \tilde A^{a\underline{\beta}]}
+\tilde A^{m\alpha}\tilde A^{n\beta} f^a{}_{mn}
+2 \tilde A^{i[\underline{\alpha}} \partial^{\underline{\beta}]}
e^a{}_i\, .
}
In addition, there exists the flux
\eq{F_\alpha&=-\partial_m \tilde A^m{}_\alpha-\tilde\partial^m A_{m\alpha
}+2D_\alpha d\, .
}

\section{Lie algebroids}
\label{app_lie}

A Lie algebroid is specified by three pieces of information:
\begin{itemize}

\item a vector bundle $E$ over a manifold $M$,

\item a bracket $[\cdot,\cdot ]_E : E \times E \rightarrow E$, and

\item a homomorphism $\rho : E \rightarrow TM$ called the anchor.

\end{itemize}
Similar to the usual Lie bracket, one  requires the bracket $[\cdot,\cdot]_E$ to satisfy a Leibniz rule.
Denoting functions by $f\in {\cal C}^{\infty}(M)$ and sections  of $E$ by $s_i$, this reads
\eq{
\label{Leibniz}
[ s_1, f s_2]_E = f \hspace{1pt}[ s_1,s_2]_E + \rho(s_1)(f) s_2  \,,
}
where $\rho(s_1)$ is a vector field which acts on $f$ as a derivation.
If  in addition the bracket $[ \cdot,\cdot]_E$ satisfies a Jacobi identity
\eq{
\label{jacobii}
\bigl[  s_1, [ s_2, s_3 ]_E \bigr]_E = \bigl[ [  s_1, s_2 ]_E , s_3 \bigr]_E +  \bigl[ s_2, [ s_1, s_3 ]_E\bigr]_E\, ,
}
then $(E,[\cdot,\cdot]_E, \rho)$ is  called
a Lie algebroid.

\smallskip

Moreover, any Lie algebroid can be
equipped with a nilpotent exterior derivative as follows
\eq{\label{algd}
	d_E\, \theta^*(s_0,\dots,s_n) =&\sum_{i=0}^n(-1)^i\,\rho(s_i)\,
		\theta^*(s_0,\dots,\hat{s_i},\dots,s_n) \\
		&+\sum_{i<j}(-1)^{i+j}\,
		\theta^*([s_i,s_j]_E,s_0,\dots,\hat{s_i},\dots,\hat{s_j},\dots,s_n)\,,
}
where $\theta^*\in\Gamma(\Lambda^nE^*)$ is the analog of an
$n$-form on the Lie algebroid and $\hat s_i$ denotes the omission of that entry.
The Jacobi identity of the bracket $[\cdot,\cdot]_E$ implies
that \eqref{algd} satisfies $(d_E)^2=0$. The anchor property and the corresponding
formula for the de~Rahm differential allow to compute
\eq{\label{diffs}
	\Big(\big(\Lambda^{n+1}\!\rho^*\big)(d_E\,\theta^*)\Big)(X_0,\dots,X_n)
	&= \big(d_E\,\theta^*\big)\big(\rho^{-1}(X_0),\dots,\rho^{-1}(X_n)\big) \\
	&= d\big((\Lambda^n\!\rho^*)(\theta^*)\big)(X_0,\dots,X_n)
}
with the dual anchor $\rho^*=(\rho^t)^{-1}$ and for sections $X_i\in \Gamma(TM)$.
The relation \eqref{diffs} describes how exact terms translate in general.

\clearpage


\bibliographystyle{utphys}
\bibliography{references}


\end{document}